\documentclass[fleqn,usenatbib]{mnras}

	\usepackage{ulem}  
	\usepackage{color} 
	\usepackage{multicol}
	\usepackage{threeparttable}
	\usepackage{booktabs}
	\usepackage{natbib}
	\usepackage{amssymb}
	\usepackage{graphicx}
	\graphicspath{{figs/}}

	\def\aap{A\& A}
	
	\def\aj{{AJ}}
	\def\apj{ApJ}
	\def\apjs{ApJS}
	\def\apjl{ApJL}
	\def\apss{Ap\& SS}
	
	\def\araa{ARA\&A}
	
	\def\mnras{{MNRAS}}

	\def\pasp{{PASP}}
	
	\def\ssr{{Space Sci. Rev.}}

	\def\ppcc{{\rm cm}^{-3}}
	\def\hii{{\sc Hii}}
	\def\prom#1{$\langle #1\rangle$}
	
	\newcommand{\Msun}{M_{\odot}}
	\newcommand{\msol}{M_{\odot}}
	\newcommand{\Myr}{{\rm Myr}}
	\newcommand{\nthr} {n_{\rm thr}}
	\newcommand{\Msink}{M_{\rm Sink}}
	\newcommand{\ergs}{{\rm erg} \, {\rm s}^{-1}}
	\newcommand{\cmcub}{{\rm ~cm}^{3}}
	\newcommand{\FUV}{F_{\rm UV}}
	\newcommand{\pc} {{\rm pc}}
	\newcommand{\K}{{\rm K}}
	
	\newcommand{\kms}{{\rm km~s}^{-1}}

	\newcommand{\MI}{M_{\rm I}}
	\newcommand{\vrms}{v_{\rm rms}}
	
	\newcommand{\ri}{R_{\rm i}}
	\newcommand{\Rinf}{R_{\rm flow}}
	\newcommand{\Linf}{L_{\rm flow}}
	\newcommand{\mG}{\mu \mrm{G}}
	
	\newcommand{\dd}{d}
	\newcommand{\eV}{{\rm ~ eV}}

	\newcommand{\beq}{\begin{equation}}
	\newcommand{\eeq}{\end{equation}}
	\newcommand{\mrm}{\mathrm}
	
	\newcommand{\Mach} {\mathcal{M}}

\title[Expansion of  H\,{\normalsize\it II} Regions in structured MCs]
            {Structure and Expansion Law of H\,{\Large\bf II} Regions 
            in structured Molecular Clouds}

\author[Zamora-Avil{\'e}s et al. ]{
        Manuel~Zamora-Avil{\'e}s,$^{1,2,3}$ 
          \thanks{E-mail:mzamora@inaoep.mx} 
        Enrique~V{\'a}zquez-Semadeni,$^2$ 
        Ricardo~F.~Gonz{\'a}lez,$^2$ 
        \newauthor Jos{\'e}~Franco,$^4$ 
        Steven N. Shore,$^{5,6,7}$ Lee W. Hartmann,$^3$ 
        \newauthor Javier Ballesteros-Paredes,$^{2}$ 
        Robi~Banerjee,$^8$ 
        Bastian~K{\"o}rtgen$^8$\\
\\
$^{1}$CONACYT-Instituto Nacional de Astrof{\'i}sica, {\'O}ptica y Electr{\'o}nica, Luis E. Erro 1, 72840 Tonantzintla, Puebla, M{\'e}xico \\
$^{2}$Instituto de Radioastronom{\'i}a y Astrof{\'i}sica,
    Universidad Nacional Aut{\'o}noma de M{\'e}xico,
    Apdo. Postal 72-3 (Xangari), Morelia,\\
    Michoc{\'a}n 58089, M{\'e}xico \\
$^{3}$Department of Astronomy, University of Michigan,  500
    Church Street, Ann Arbor, MI 48105, USA \\
$^4$Instituto de Astronom{\'i}a, Universidad Nacional Aut{\'o}noma de M{\'e}xico, 
    AP 70-264, CDMX, C.P. 04510, M{\'e}xico \\
$^5$Dipartimento di Fisica ``Enrico Fermi'',  Universita' di Pisa, Italy \\
$^6$INFN Pisa, Largo B. Pontecorvo, 56127, PI, Italy \\
$^7$Astronomical Institute, Charles University in Prague, 
    V Hole\v{s}ovi\v{c}k\'ach 2, 180 00, Praha 8, Czech Republic\\
$^8$Hamburger Sternwarte, Universit{\"a}t Hamburg, Gojenbergsweg 112, 
    21029 Hamburg, Germany
}

\date{Accepted XXX. Received YYY; in original form ZZZ}

\pubyear{2018}

\begin{document}

\label{firstpage}
\pagerange{\pageref{firstpage}--\pageref{lastpage}}
\maketitle

\begin{abstract}
We present radiation-magnetohydrodynamic simulations aimed at studying evolutionary properties of H\,{\normalsize II} regions in turbulent, magnetised, and collapsing molecular clouds formed by converging flows in the warm neutral medium. We focus on the structure, dynamics and expansion laws of these regions. Once a massive star forms in our highly structured clouds, its ionising radiation eventually stops the accretion (through filaments) toward the massive star-forming regions. The new over-pressured H\,{\normalsize II} regions push away the dense gas, thus disrupting the more massive collapse centres. Also, because of the complex density structure in the cloud, the H\,{\normalsize II} regions expand in a hybrid manner: they virtually do not expand toward the densest regions (cores), while they expand according to the classical analytical result towards the rest of the cloud, and in an accelerated way, as a blister region, towards the diffuse medium. Thus, the ionised regions grow anisotropically, and the ionising stars generally appear off-centre of the regions. Finally, we find that the hypotheses assumed in standard H\,{\normalsize II}-region expansion models (fully embedded region, blister-type, or expansion in a density gradient) apply simultaneously in different parts of our simulated H\,{\normalsize II} regions, producing a net expansion law ($R \propto t^\alpha$, with $\alpha$ in the range of 0.93-1.47 and a mean value of $1.2 \pm 0.17$) that differs from any of those of the standard models.
\end{abstract}



\section{Introduction} \label{sec:intro}

Massive stars play a key role in the evolution of galaxies. Through a
combination of massive outflows, expanding \hii\ regions, and supernova
explosions, they shape and provide an important input of energy to the
interstellar medium \citep[ISM; e.g.,][]{MacLow-Klessen04}. In
particular, they erode and disperse their parent molecular clouds (MCs),
directly affecting the star formation activity within the clouds
\citep[see, e.g.,][for recent reviews]{Krumholz+14,VS15}.

It is generally thought that the
negative feedback\footnote{By {\it negative feedback} we refer to
the suppression of star formation by erosion of the dense regions where
massive stars form. Similarly, we will refer to the promotion of
star formation by expanding
\hii\ regions as {\it positive feedback}, as in the classical {\it collect and collapse} scenario \citep{Elmegreen-Lada77}.}
through blister-type \hii\ regions \citep[or ``champagne''
flows;][]{FTB90} is efficient in eroding and dispersing MCs on
timescales of few tens of Myr \citep{Blitz-Shu80, Matzner02}. Idealised
analytical \citep[e.g.,][hereafter, we will refer to the latter work 
as FST94]{Whitworth79, Franco+94} and numerical
\citep[e.g. ][]{Bodenheimer+79, Tenorio-Tagle79} works have shown that
blister \hii\ regions are also able to reduce the star formation
efficiency (SFE) of MCs to the low observed values of $\lesssim 10$\%
\citep[e.g.,][]{Myers+86}. However, all these simplified models without
self-gravity, with plane-parallel geometry and/or uniform density fields
are far from the complex morphology and dynamics observed in MCs
\citep[see, e.g.,][for a recent review]{Andre+14}, which includes
turbulence, magnetic fields and anisotropic and hierarchical collapse,
leading to the formation of filamentary structures that funnel gas to
the star-forming sites \citep[e.g.,][although see Matzner \& Jumper 2015
for an exception] {GV14, Smith+16, VS+17}.

Furthermore, the dynamics (free-fall motions or accretion) and the
high-density environment of the birthplaces of massive stars could
strongly attenuate the disruptive effect of the massive stars
\citep{Yorke+89, Dale+05, Peters+10}. This picture gets more
complicated if we take into account that MCs could be in global
collapse, as recent and growing evidence shows \citep[see,
e.g.,][]{HBB01, BH04, HB07,PHA07, VS+07,VS+09, Galvan+09, Schneider+10,
Csengeri+11, BP+11, Hartmann+12, BP+15,Peretto+14, Carmen+17}.

In a medium with power law density stratification $r^{-p}$, with $p > 3/2$, 
\citet{FTB90} showed analytically that \hii\ regions expand in an
accelerated way. \cite{Arthur+06}, using numerical simulations of
\hii\ regions expanding in a stratified medium that decreases
exponentially, showed that a very weak shock
develops toward the densest part of the cloud. This result is crucial
in interpreting analytical models in the literature (see
App. \ref{ap:models}).

At scales of MCs, recent numerical simulations (regardless of the
initial conditions or setup) of highly-structured clouds have shown that
negative feedback is able to reduce the SFE of MCs to values consistent
with the observations \citep[e.g.,][]{Dale+05, Walch+12,Colin+13, Geen+15}. The
effect of the ionising feedback depends strongly on the clouds' masses
and sizes, being weaker for more massive clouds
\citep{Dale+12}. However, simulations investigating this effect on
clouds of various masses generally tend to assume idealised, spherical
initial mass distributions, which are not necessarily realistic, since
it is known that clouds tend to be sheet-like and filamentary
\citep[e.g.,] []{Bally+87, Andre+14}. Thus, it is important to
investigate the effect of stellar feedback in realistically-shaped
clouds, whose morphology is dictated self-consistently by their
evolution since their formation.

In contrast, several
observational studies have attempted to infer empirical correlations between
physical properties of \hii\ regions, such as density and
size. For example, \citet{Hunt+09} studied extragalactic \hii\
regions and compiled data from both Galactic
\citep{Garay-Lizano99, KK01, Martinez-Hernandez+2003, Dopita+06} and
extragalactic \citep{Kennicutt84,GG07} samples, concluding that the
entire sample follows a size ($\ri$) {\it versus} ionised gas density
($n_i$) trend of the form $n_i \propto \ri^{-1}$, although with a
considerable dispersion. Interestingly, combining this relation and the
Larson correlation ($n_{\rm H_2} \propto R_{\rm H_2}^{-1.1}$; where
$n_{\rm H_2}$ and $R_{\rm H_2}$ are the density and size of a given MC)
these authors suggest that the star formation is a scale-free
process. However, there have been claims in the literature that the
Larson density-size correlation is only the result of a selection effect
due to the criteria used for defining a MC and their rapidly
decaying column density PDFs (\cite{BP+12}, see also
\cite{Kegel89, Scalo90, VS+97, BP-MacLow02, Heyer+09, Camacho+16}
for other possibilities).

In order to study the  evolutionary properties of
individual \hii\ regions embedded in realistically structured MCs, we present in this contribution radiation-magnetohydrodynamic (RMHD)
simulations of MCs formed by converging flows, that
evolve self-consistently, from their formation to their destruction by
ionising radiation. Because of this self-consistent evolution, the
clouds also have a realistic spatial
structure, which, rather than spherical, is closer to being
sheet-like, and highly inhomogeneous. 
In this work, we also study 
the time evolution of the size-density relation of the ionised gas.

We organise the paper as
follows. In \S \ref{sec:NS-model}, we describe the numerical model. In
\S \ref{sec:NS-results} and \ref{sec:structure} we present our results, which are then
discussed in \S \ref{sec:discusion}. Finally, the summary and
conclusions are presented in \S \ref{sec:conclusions}.

\section{The numerical model} \label{sec:NS-model}

With the goal of investigating the effect of magnetic fields
in the formation and evolution of MCs, in \citet[][hereafter Paper
I]{ZA-VS+17} we presented three-dimensional, self
gravitating, MHD simulations of MCs formed by two WNM colliding flows,
including heating and cooling processes. These simulations were
carried out using the Eulerian adaptive mesh refinement {\tt FLASH}
(v2.5) code \citep{Fryxell+00}.

In this contribution we consider one of those models
(labelled B3), in which the magnitude of the
magnetic field (initially uniform) is $3 \mu$G in our WNM-like initial
conditions. This value is consistent with the observed mean value magnetic
field of the uniform component in the Galaxy \citep{Beck01}. In
addition, we present an additional simulation
including radiative transfer, in order to model the effects of UV
feedback from massive stars. For this, we use the radiation scheme
introduced by \citet{Rijkhorst06} and improved by
\citet{Peters+10}.\footnote{See also \citet{Buntemeyer+16}.} This implementation has successfully passed several
tests. It accurately follows the velocity propagation for R-type
\citep[in a cosmological context; ][]{Iliev+06} and D-type
\citep[corresponding to the analytical Spitzer solution; ][]{Peters+10}
ionisation fronts. In this section, we focus on the description of the
radiation module and for further information we refer the reader to
Paper I.

\subsection{Sink particles and refinement criterion} \label{subsec:sinks}

As in Paper I, we use the so called {\it constant mass} criterion in order to follow the evolution of high density regions. According to this criterion, the grid size scales with density as $\Delta x \propto \rho^{-1/3}$, and so we refine once the cell density is eight times larger than in the previous level to guarantee that the mass of each cell is preserved.
At the maximum level of refinement a sink particle can be formed when the density in this cell exceeds a threshold number density, $\nthr \simeq 4.2 \times 10^{6} \, \ppcc$, among other standard sink-formation tests. Once the sink is formed, it can accrete mass from its surroundings \citep[][]{Federrath+10}.

Note that this criterion is not standard given that it does not fulfil the Jeans criterion \citep{Truelove+97}, which states that artificial fragmentation can be avoided if the Jeans length is resolved with at least four grid cells. 
However, in Paper I we showed that the only effect of using the Jeans
criterion rather than the {\it constant mass} one is a
slight delay in the onset of star formation, leaving unchanged the sink
mass distribution, which determines the intensity of the UV feedback
sources. The reason we chose the {\it constant mass} criterion over the Jeans one is that the refinement is concentrated in the densest gas, which allows us to speed up the calculations.


\subsection{Subgrid Star Formation prescription} \label{subsec:sf-recipe}

Given the size of our numerical box and the maximum resolution we can achieve, the sink particles rapidly reach hundreds of solar masses via accretion, and therefore we must not treat them as single stars but rather as groups of stars. Thus, we assume a standard initial mass function (IMF) and we estimate the most massive star that the sink can host, which dominates its UV flux. The sink radiates according this flux. We use a \citet{Kroupa01}-type IMF, which reads
\begin{equation}
\chi (m) \propto m^{-\alpha_i},
\end{equation}
where $\alpha$ is a piecewise constant, and $\dd N=\chi (m) \dd m$ is the number of single stars in the mass 
interval $m$ to $m+\dd m$. We normalise this IMF as
\beq \label{eq:NS-NORM}
\int_{0.01 \msol}^{60 \msol}\, m \, \chi (m) \dd m = \Msink,
\eeq
where $M_{\rm sink}$ represents the individual mass of the sink particles. 
We take the standard lower and upper limits of $0.01$ and $60 \, \msol$, respectively. We then integrate $\dd N$ over bins of $1 \, \Msun$ to obtain the number of stars, $\Delta N$, in each mass bin. The centre of the last bin ($m_*$) satisfying $\Delta N \gtrsim 1$ is taken as the mass of the most massive star that the sink can host.

Finally, as described in \cite{Peters09, Peters+10}, from zero-age main sequence (ZAMS) models \citep[assuming solar metallicity;][]{Paxton04} we assign an UV flux (of photons with energy higher than 13.6 eV) to the most massive star, $m_*$, and we assume that this star dominates the emission of ionising photons from the sink.\footnote{We do not take into account that massive stars have frequently a companion of similar mass so this would have an effect on the \hii\ region evolution.} In order to save computational time, we assume that only sinks containing stars with masses $\gtrsim 8 \, \Msun$\footnote{which correspond to $\FUV \simeq 10^{46.5} \, {\rm s}^{-1}$ for a cluster of $\simeq 83 \, \Msun$.} emit ionising radiation, since stars with lower masses ($< 8 \, \Msun$) do not emit significant amounts of photoionising photons. We allow the massive stars to radiate for $5 \, \Myr$.\footnote{Note, however, that this period does not affect our results since we are interested on the properties of \hii\ regions at the initial stages of evolution.}

\subsection{Feedback prescription}

We use an adapted version of the hybrid characteristic ray-tracing module in the {\tt FLASH} code \citep{Rijkhorst06,Peters+10}. The method can be summarised as follows. To calculate the flux of ionising photons arriving at each cell, the column density is calculated by interpolation (grid mapping) along rays from the point sources to every cell. Then the ionisation fractions and temperature can be computed through an iterative process (under the assumption of radiative equilibrium), taking advantage of the analytic solution to the rate equation for the ionisation fractions. Furthermore, the heating/cooling can be iterated to convergence (see Sec. \ref{sec:NS-heating-cooling}), so that the only restriction on the time-step comes from the MHD module. The MHD and ionisation calculations are coupled through operator splitting \citep[see also][]{Frank-Mellema94, Mellema-Lundqvist02}. Throughout this paper, we assume solar metallicity/abundances. For a detailed description about the feedback implementation in FLASH code we refer the reader to \citet{Peters09, Peters+10}. 

\subsection{Heating and Cooling} \label{sec:NS-heating-cooling}

Following \citet{Krumholz+07b}, we calculate the heating and cooling
rates by breaking them into heating and cooling associated with
ionisation of hydrogen atoms by point sources (the sink
particles), and other relevant sources of cooling and heating. For the former, the
photoionisation rate is \citep{Osterbrock89}
\beq
\Gamma_{\mrm{ph}} = n_{\mrm{HI}}  \int_{\nu_{\mrm{T}}}^\infty
\frac{4 \pi J_\nu}{h \nu} \sigma_\nu h (\nu-\nu_{\mrm{T}}) \dd \nu,
\eeq 
where $n_{\mrm{HI}}$ is the number density of atomic hydrogen, $\nu$ is the frequency and $\nu_{\mrm{T}}$ is the ionisation threshold frequency (at $13.6 \, \eV$), $\sigma_\nu$ is the absorption cross section of atomic hydrogen, and $h$ is the Planck constant. The specific mean intensity, $J_\nu$, of a point source/star of radius $r_{\mrm{star}}$ and effective temperature $T_{\mrm{star}}$ (assuming a blackbody spectrum) is given by \citep{Rijkhorst06,Peters+10}
\beq \label{eq:NS-Jnur}
J_\nu(r)=\Big( \frac{r_{\mrm{star}}}{r} \Big)^2
\frac{1}{2c^2}
\frac{h \nu^3}{\exp(h \nu / k_{\mrm{B}} T_{\mrm{star}})-1}
\exp(-\tau_\nu(r)).
\eeq
with $\tau_\nu(r)$ the optical depth at position $r$ computed directly from the column density, $N(r)$. We also take into account the dust heating term ($\Gamma_{\rm d}$) by non-ionising radiation \citep[e.g.,][]{Krumholz+07, Peters+10}.
%
%
To counterbalance the photoionisation heating rate, $\Gamma_{\mrm{ph}}$, we consider the co\-lli\-sio\-nal cooling (ions-electrons), $\Lambda_{\rm col}$, which is the main mechanism for energy loss in partially ionised gas \citep[see, e.g.,][]{Dalgarno-McCray72}. 

For heating and cooling that are not directly due to
ionisation from the sink particles, we use the
analytic fits by \citet{KI02}\footnote{See also \cite{VS+07} for
corrections to typographical errors in the original source paper}. 
for the heating ($\Gamma_{\rm KI}$) and cooling ($\Lambda_{\rm KI}$)
functions,

\beq
\Gamma_{\rm KI} = 2.0 \times 10^{-26}  \, \ergs
\eeq
\beq
\frac{\Lambda_{\rm KI}(T)}{\Gamma_{\rm KI}} = 10^7 \exp{\frac{-1.184 \times 10^5}{T+1000}}+1.4 \times 10^{-2} \sqrt{T} \exp{\frac{-92}{T}} \cmcub,
\eeq
which are based on the thermal and chemical calculations
considered by \citet{Wolfire+95, KI00}, including photoelectric heating from small grains and PAHs, heating and ionisation by X-rays, cosmic rays, and H$_2$ formation/destruction. Cooling processes include atomic line emission from C\,{\normalsize II}, O\,{\normalsize I}, hydrogen Ly$\alpha$, 
rotation/vibration line cooling from  CO and H$_2$, and atomic and molecular collisions with dust grains.  

Thus, the net heating and cooling rates are
\beq
\Gamma = \Gamma_{\mrm{ph}} +  \Gamma_{\rm d}  +  \Gamma_{\rm KI},  \qquad 
\Lambda = n_e n_{\mrm{HII}}\Lambda_{\rm col} +  n^2_{\mrm{HI}}\Lambda_{\rm KI}
\eeq
where $n_{\mrm{HI}}$, $n_e$, and $n_{\mrm{HII}}$ refers to the number
density of neutral gas, electrons, and ionised gas, respectively.

\begin{figure}
     \includegraphics[width=1\hsize]{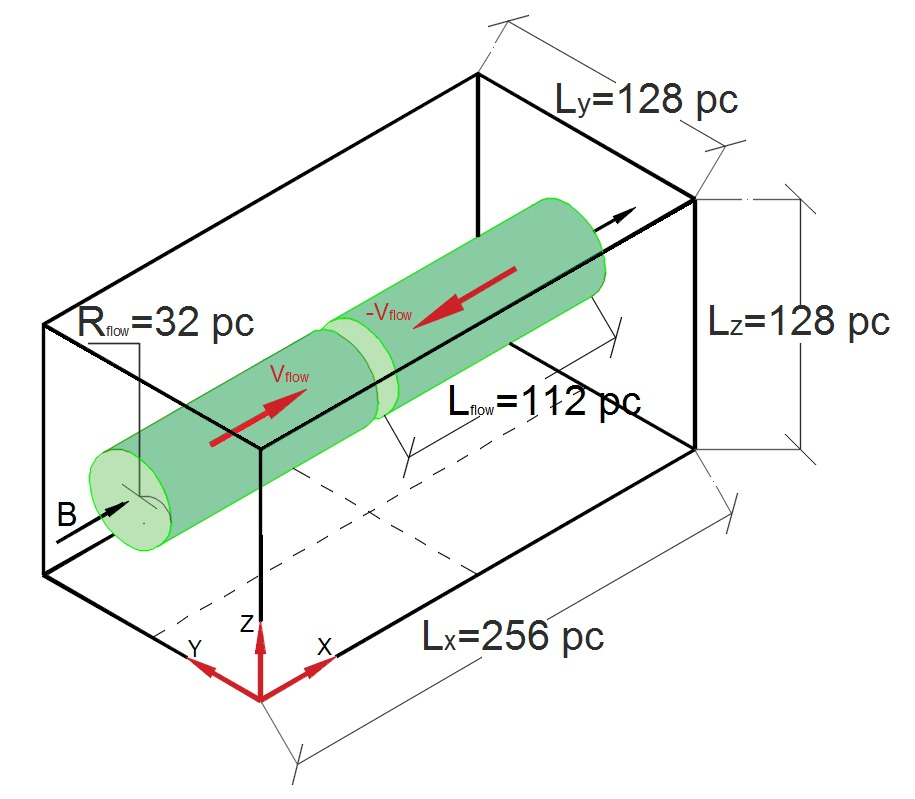}
     \caption{Sketch illustrating the initial conditions used in this work, which consist of two cylindrical streams colliding at the centre of the numerical box. 
     Figure adapted from \citet{Bastian15}.}
    \label{fig:ics}
\end{figure}

\subsection{Initial conditions} \label{subsec:NS-initial-conditions}

We use the setup pictured in Fig. \ref{fig:ics} for the initial 
conditions, which are as follows. 
The numerical periodic box, of sizes $L_x=256 \, \pc$ and $L_y = L_z
=128 \, \pc$, is initially filled with warm neutral gas at uniform
density of $2 \, \ppcc$ and constant temperature of $1450 \,
\K$.\footnote{This temperature corresponds to the thermal equilibrium
and implies an isothermal sound speed of $c_{\rm s} \simeq 3.1 \, \kms$.}  

We impose an initial background velocity field, which corresponds to a
moderate
turbulence with a power spectrum of $k^{-2}$ and Mach number of
$\Mach_{\rm rms} \simeq 0.7$, whose main role is to trigger instabilities in the
converging-flow we set up on top of this random field. This setup
consists of two cylindrical streams entirely contained in the numerical domain (see Fig. \ref{fig:ics}), each of radius $\Rinf=32 \, \pc$
and length $\Linf = 112 \,
\pc$, moving in opposite directions at a moderately supersonic velocity
of $v_{\rm flow}=7.5 \,  \kms$ in the $x$-direction. Thus, the
inflow Mach number is $\Mach_{\rm flow}=2.42$ and the corresponding
dynamical time is $t_{\rm dyn}=\Linf /v_{\rm flow}=14.6$ Myr.
Note that the gas mass in the whole box is $\sim 2.6 \times 10^5 \,
\Msun$, whereas the mass contained in the cylinders is $\sim 4.5 \times
10^4 \, \Msun$ (assuming a mean molecular weight $\mu = 1.27$). 

The numerical box is permeated {\rm by} a uniform magnetic field of $3 \, \mG$ along the $x$-direction.
Thus, the corresponding mass-\-to-\-flux ratio in the cylinders is $1.59$ times the critical value, so that the cloud formed by the colliding flows eventually will become mag\-ne\-ti\-cally supercritical once it accrete enough mass. The initial plasma beta parameter (i.e., the thermal to magnetic pressure ratio) is $\beta_{\rm th} \equiv P_{\rm th} /  P_{\rm mag} \simeq 1.1$ and the Alfv\'en mach number $\Mach_{\rm A} = 1.3$. We achieve a maximum resolution of $\Delta = 0.03 \, \pc$ in all the three dimensions. 

With this setup we ran model B3 in Paper I ({\it Feedback-Off} model
hereafter). In this work, we restart this simulation\footnote{We used this model since its initial magnetic field ($3 \, \mu G$) is more consistent with the uniform component strength observed in the Galaxy \citep{Beck01}.}
to include
ionising feedback from the point when the first massive star appears
($t \sim 12.8 \, \Myr$; model {\it Feedback-On} hereafter) and let it
evolve for $\sim$6 Myr more. This period is long enough to study the
evolution of individual \hii\ regions as well as the global feedback
effects on the parent MC.

\begin{figure*}
     \includegraphics[width=1\hsize]{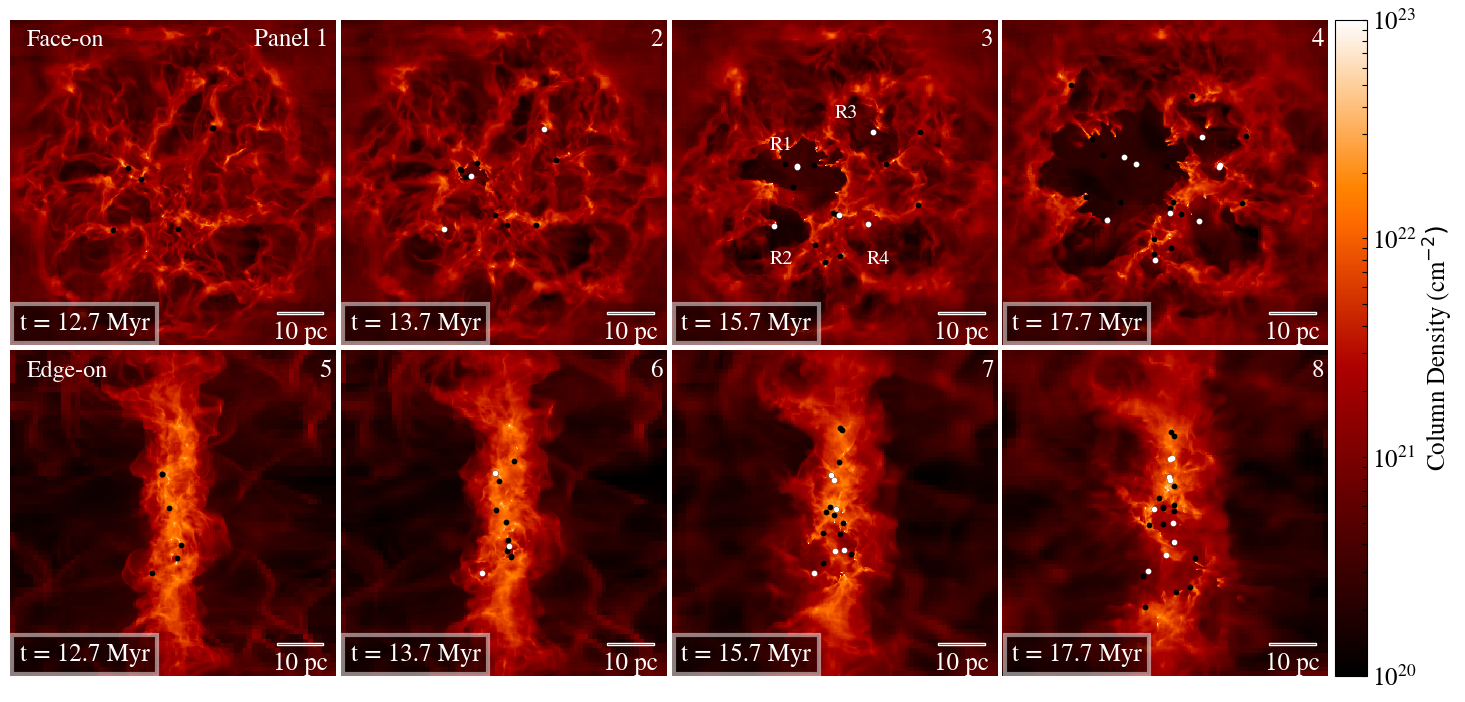}
     \caption{Column density maps in evolutionary sequence for the
``central cloud'' in the simulation with feedback (Feedback-On model). The
upper  and  lower panels show face-on and edge-on views,
respectively. The dots represent the projected position of the sink
particles, where the white ones host 
massive stars. In panel 3
we annotate the name of the four \hii\ regions we analyse in
Sec. \ref{subsec:hii-regions}. The projections correspond to the 80 pc
central sub-box. See an animation of this figure in the supplementary material.} 
    \label{fig:cloud-evolution}
\end{figure*}

\begin{figure*}
	\centering
	\begin{minipage}{\textwidth}
		\centering
		\includegraphics[width=0.96\linewidth]{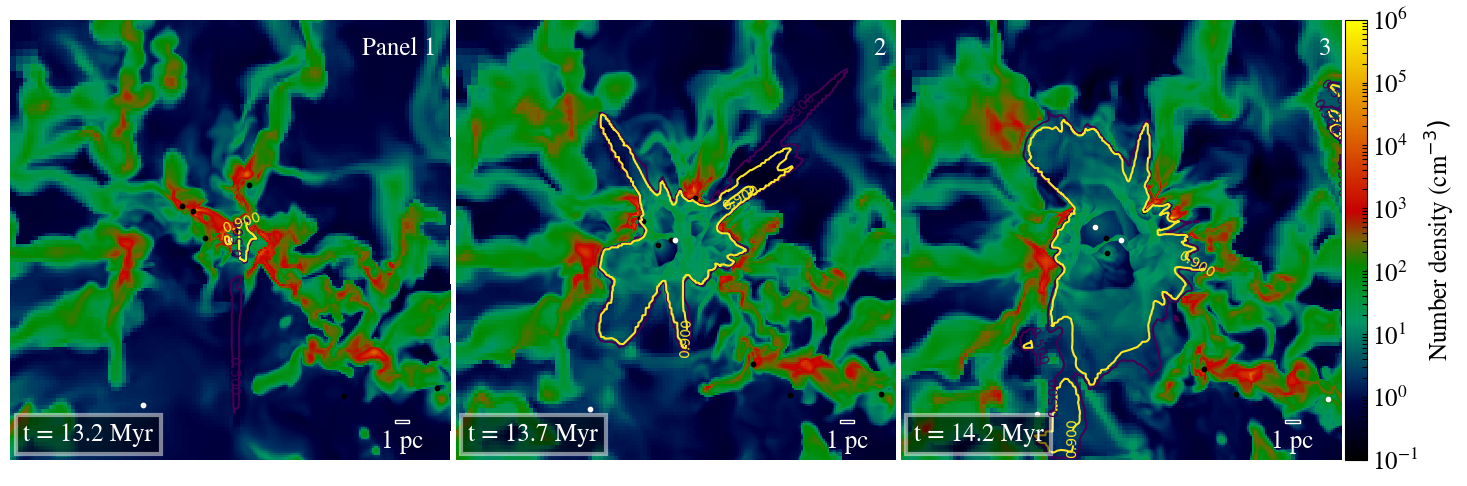}
	\end{minipage}\hfill
    	\begin{minipage}{\textwidth}
		\centering
		\includegraphics[width=0.96\linewidth]{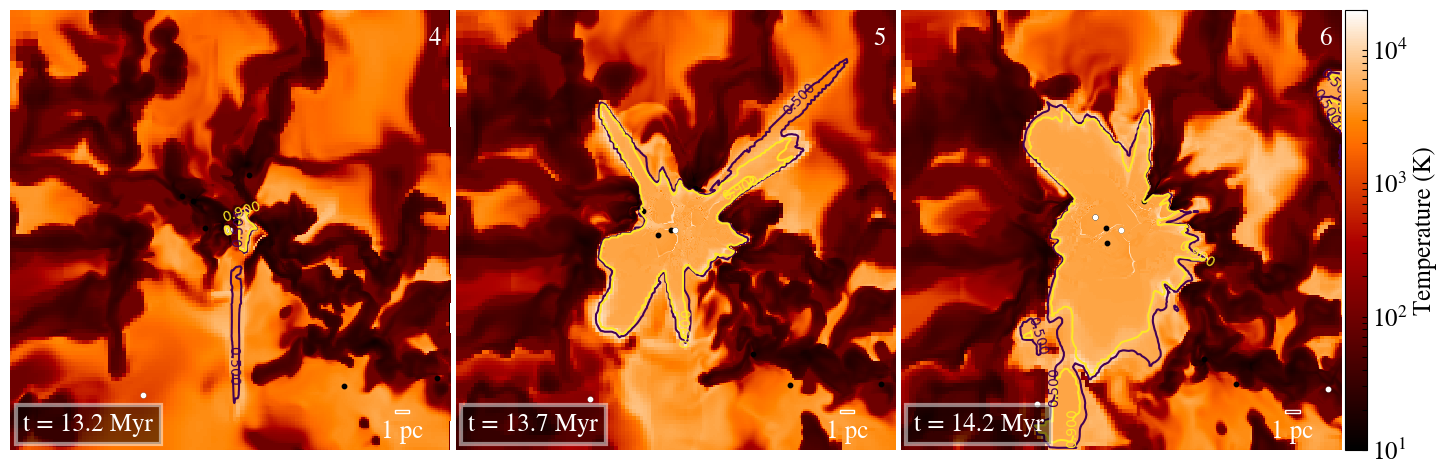}
	\end{minipage}\hfill
    \begin{minipage}{\textwidth}
		\centering
		\includegraphics[width=0.96\linewidth]{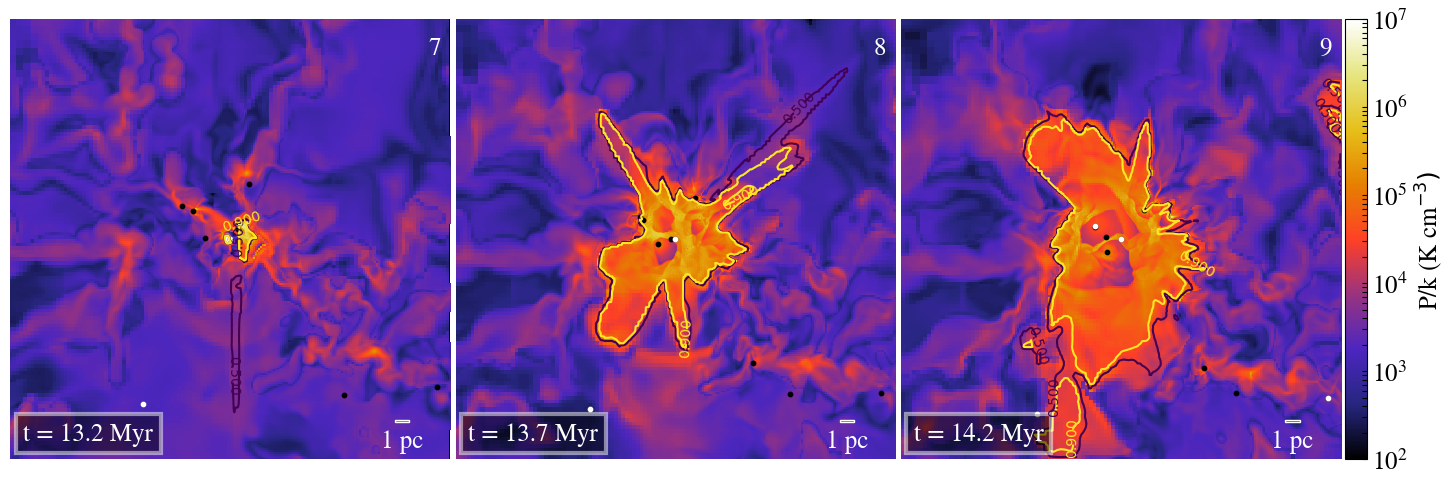}
	\end{minipage}\hfill
	\begin{minipage}{\textwidth}
		\centering
		\includegraphics[width=0.96\linewidth]{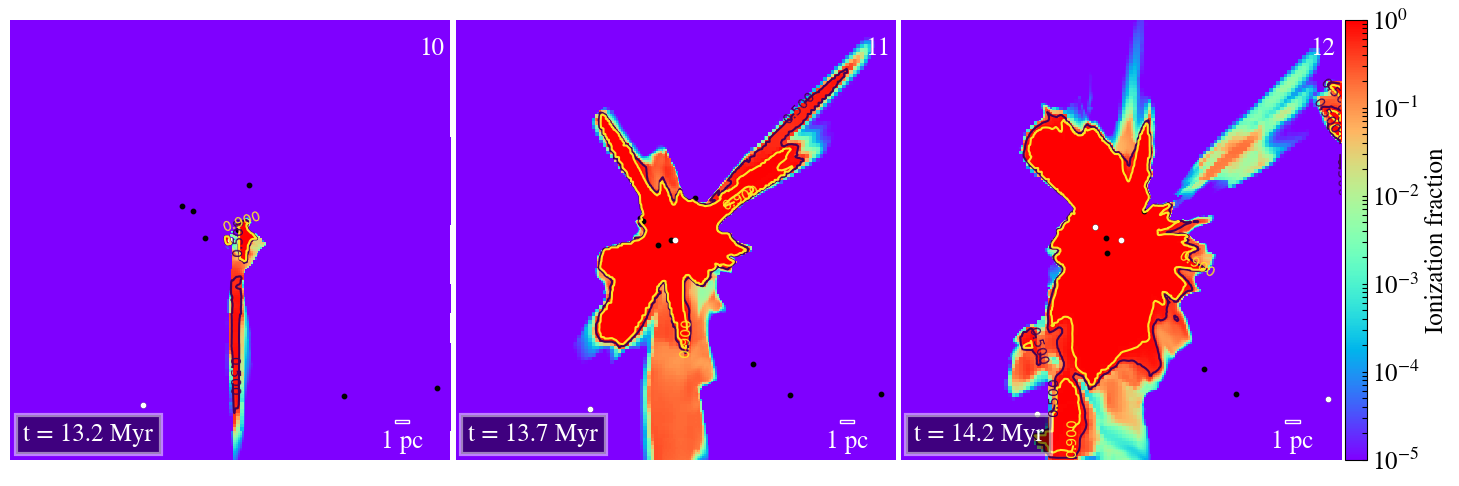}
	\end{minipage}
		\caption{Slices of the number density, temperature,
pressure, and ionisation fraction (from top to bottom rows) in the
$y-z$ plane (face-on view) for the first \hii\ region (R1) at
different times. Each slice is 30 pc wide and is centred at the
position of the massive star. The contours in all panels delineate ionisation
fractions of 0.5 (purple) and 0.9 (yellow).} 
  \label{fig:slc_x}
\end{figure*}

\begin{figure*}
	\centering
	\begin{minipage}{\textwidth}
		\centering
		\includegraphics[width=0.96\linewidth]{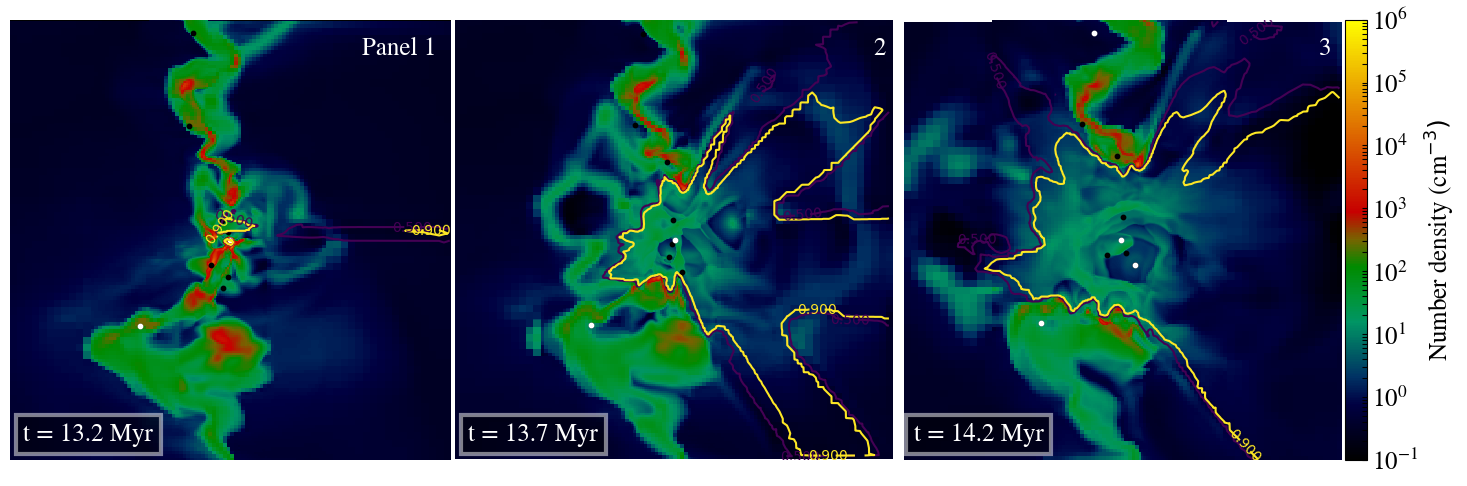}
	\end{minipage}\hfill
    	\begin{minipage}{\textwidth}
		\centering
		\includegraphics[width=0.96\linewidth]{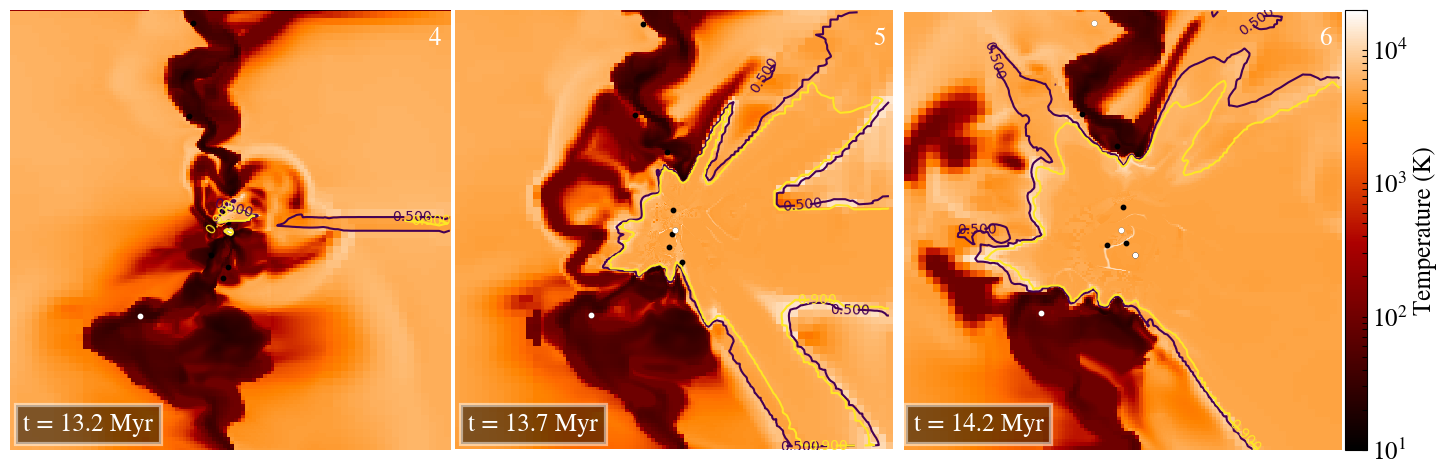}
	\end{minipage}\hfill
    \begin{minipage}{\textwidth}
		\centering
		\includegraphics[width=0.96\linewidth]{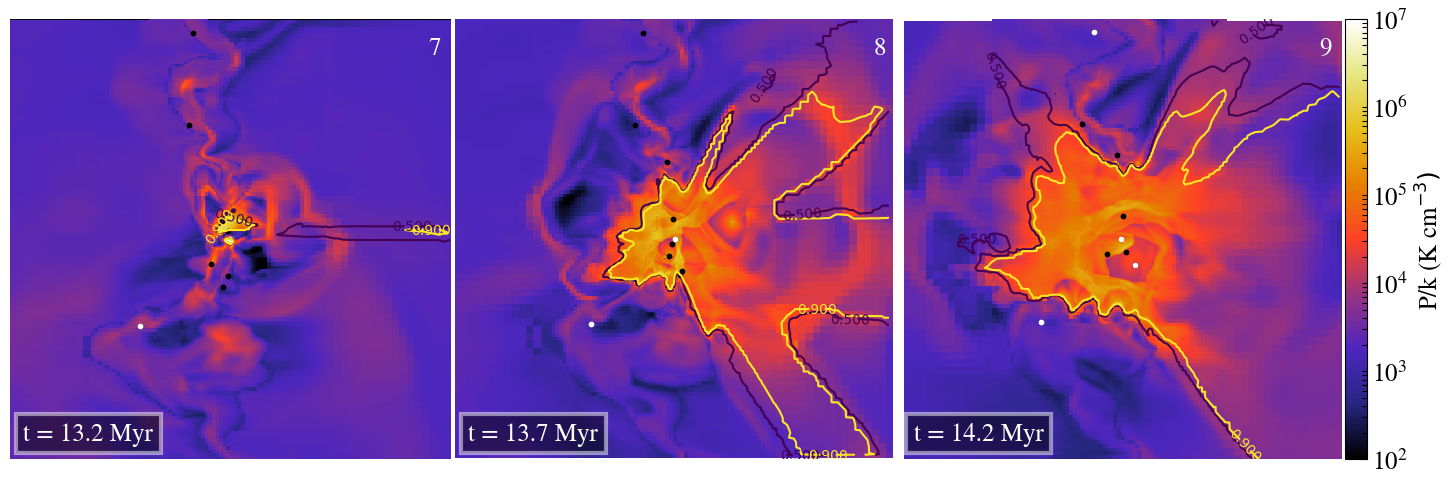}
	\end{minipage}\hfill
	\begin{minipage}{\textwidth}
		\centering
		\includegraphics[width=0.96\linewidth]{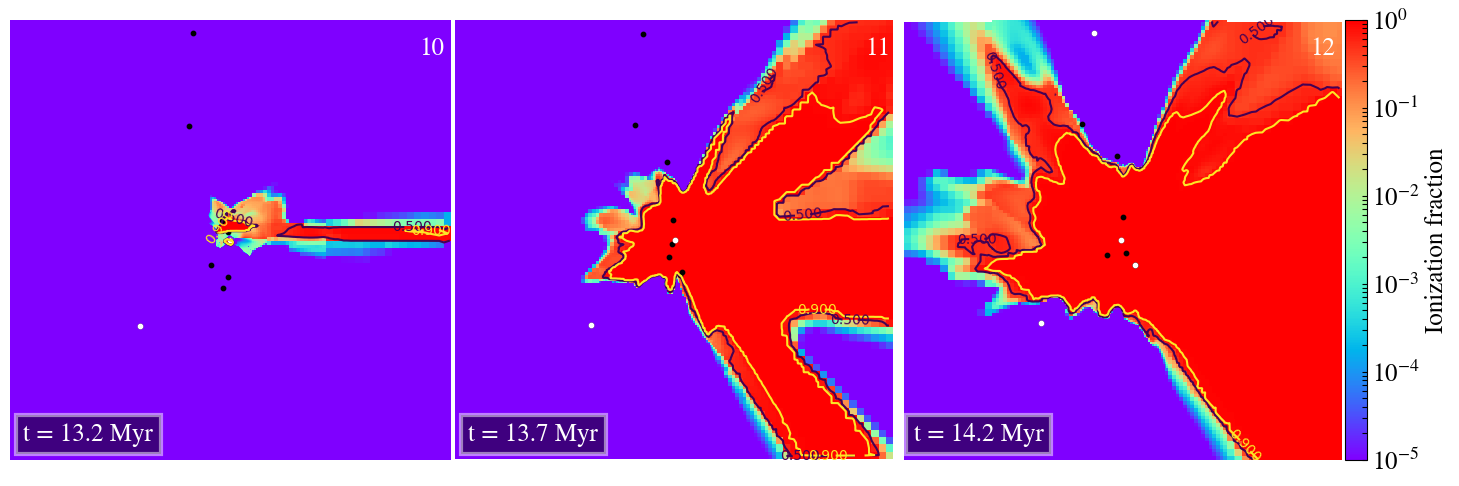}
	\end{minipage}
		\caption{Same as Fig. \ref{fig:slc_x} but in the $x-y$
plane (edge-on view).} 
  \label{fig:slc_z}
\end{figure*}

\section{Results} \label{sec:NS-results}

Henceforth, we focus our discussion on the simulation with feedback (model Feedback-On). For a detailed description about the formation and evolution of the control simulation (model Feedback-Off) we refer the reader to Paper I.

\begin{figure*}
	\centering
	\begin{minipage}{\textwidth}
		\centering
		\includegraphics[width=\linewidth]{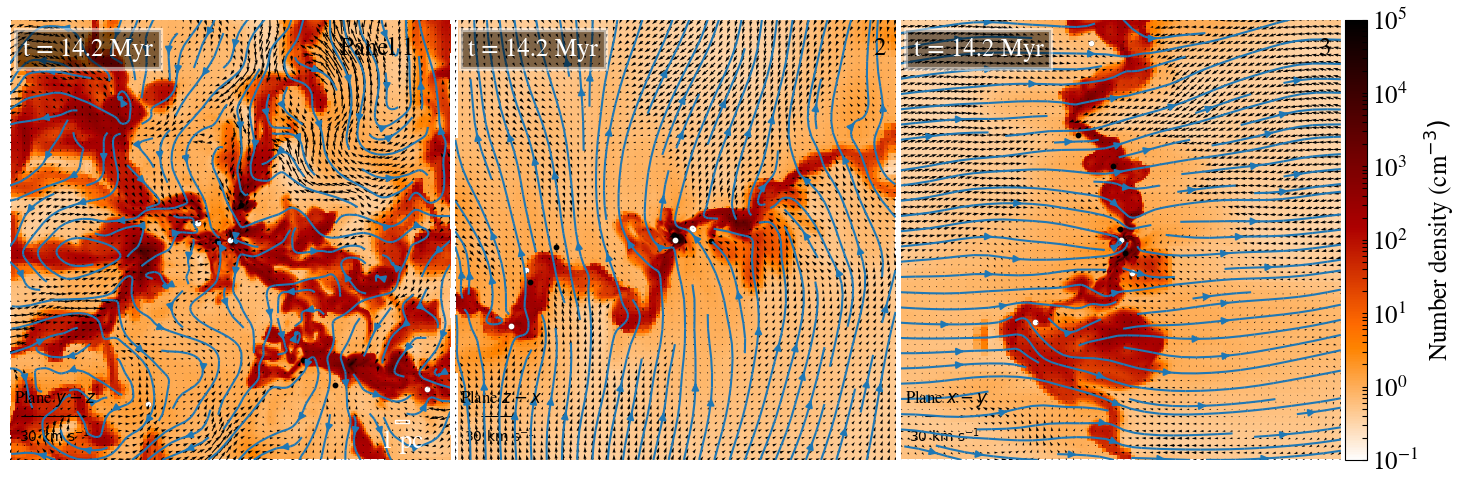}
	\end{minipage}\hfill
    	\begin{minipage}{\textwidth}
		\centering
		\includegraphics[width=\linewidth]{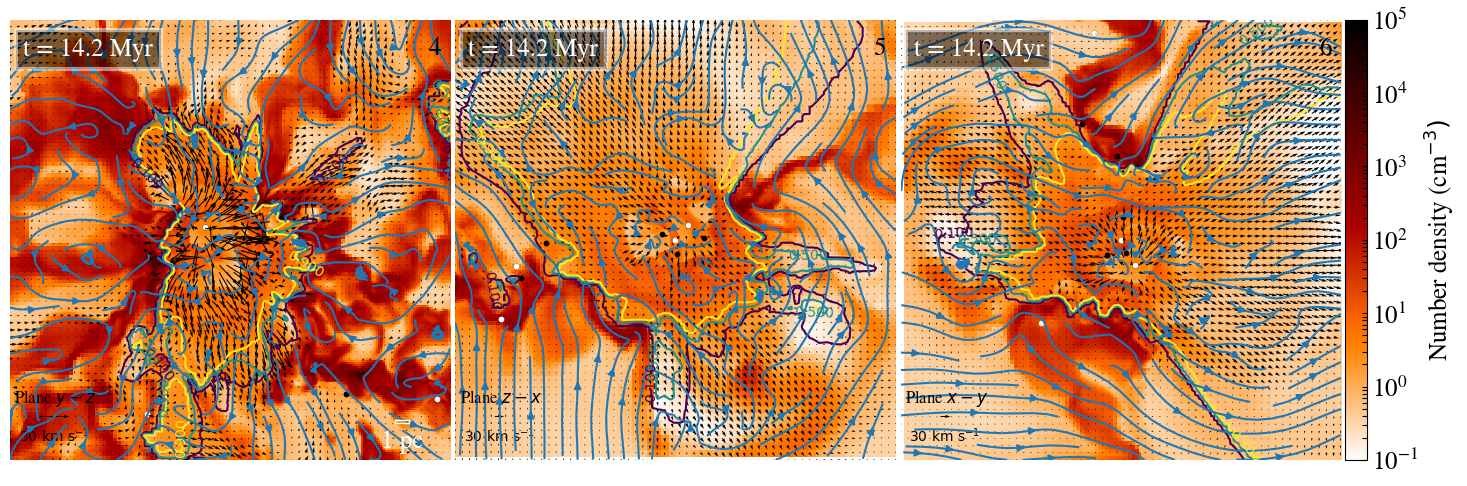}
	\end{minipage}
		\caption{Number density slices of the region ``R1" at $t=14.2 \, \Myr$ in the three main projections, $x$, $y$, and $z$ (left, middle, and right panels, respectively). The upper/lower panels correspond to the simulation without/with feedback (i.e., Feedback-Off/Feedback-On models). The black arrows represent the projected velocity field (the vector scale is in the bottom left corner). Blue lines represent magnetic field streamlines. In the lower panels (Feedback-On model), the yellow, light-blue, and purple contours correspond to ionising fractions of 0.9, 0.5, and 0.1, respectively.}
  \label{fig:vel-mag}
\end{figure*}

\subsection{Global evolution}

In general, the clouds formed in
our colliding flows simulations start as a
thin cylindrical sheet of cold atomic gas produced by the thermal
instability \citep[see, e.g.,][]{HP99,KI00,KI02,WF00, VS+06}. The
inflows naturally inject turbulence to the cloud through
various dynamical instabilities \citep[see,
e.g.,][]{Hunter+86,Vishniac94, KI02, Heitsch+05, VS+06}. The cloud
continues accumulating mass via accretion and eventually 
becomes gravitationally unstable and 
begins to contract gravitationally, entering a
regime of hierarchical, multi-scale collapse
\citep{HBB01,HH08, VS+09, VS+19}. 
Some time after that ($t \sim 11.6 \,
\Myr$), star formation begins in the densest fragments
(clumps), while these fragments continue to fall towards the global
centre of mass. The first radiating sink forms at $t \sim 11.8 \, \Myr$ (with a mass of $\sim 3.5 \, \Msun$) and starts to radiate at $t \sim 12.8 \, \Myr$ once it accretes enough mass to host a massive star, according to our subgrid SF prescription (see \ref{subsec:sf-recipe}). 

In Fig. \ref{fig:cloud-evolution} we show the column density structure
of the simulation with feedback in face/edge-on projections at
different time steps (see the corresponding movie in the supplementary material). 
Note the highly complex filamentary structure of
the dense gas (especially at $t \sim 12.7 \, \Myr$), which is a common
feature in observed molecular clouds \citep[see, e.g. ][and references
therein]{Andre+14} and simulations
\citep[e.g,][]{GV14,Smith+14,ZA-BP-LH17}. At 
later times ($t \sim 15.7 \Myr$) the effect of ionising feedback is
evident and we can see the features typical of \hii\ regions,
such as pillars, elephant trunks, and champagne flows \citep[see,
e.g.,][]{Hester+96}. In Figs.\ \ref{fig:slc_x} and
\ref{fig:slc_z} we show slices of the number density, temperature, pressure, and
ionisation fraction for the first \hii\ region that appears
(region labeled ``R1" in panel 3 of Fig. \ref{fig:cloud-evolution}) in two different 
projections and for three different early times ($t=13.2$, 13.7, and 14.2
Myr). It is worth noting that the HII regions
in these filamentary structures immersed in a WNM are far from
spherical and properties such as size are strongly
projection-dependent. 

In Fig. \ref{fig:vel-mag} we show 30-pc slices at $t=14.2$ Myr centred
at the position of the first massive star that appears for the
simulations without and with feedback (upper and lower panels,
respectively). The left, middle and right panels correspond to slices in
planes $y-z$, $z-x$, and $x-y$, respectively. The first feature worth
noting in the simulation without feedback (upper panels) is that the
filament containing the massive star is perpendicular to the magnetic
field lines \citep[see blue streamlines in panels 2 and 3; see also][]{Gomez+18}. This is probably 
a consequence of the initial conditions, however, it could be also the result
of the magnetic field being oriented by the accretion flow from the
cloud onto the filament \citep[e.g.,][]{ZA-BP-LH17}, rather than the
field guiding the flow, as such morphology is often interpreted.

In contrast, in the simulation with feedback, the ionised gas is
over-pressured and escapes in a champagne flow to the WNM at
velocities of tens of $\kms$ (see panels 5 and 6 in
Fig. \ref{fig:vel-mag}). Note that in the nearest 10 pc around
the massive star, the magnetic field lines are highly 
disorganised. However, beyond
that distance, the magnetic field is quite
ordered and tends to be aligned with the velocity
field of the outflowing gas.
This outflowing material is mostly
composed of fully ionised gas, as the ionisation fraction contours
show in panels 4-6 of Fig. \ref{fig:vel-mag} (purple, light-blue, and
yellow contours denote ionisation fractions of 0.1, 0.5, and 0.9,
respectively). Although we have been discussing mostly the
morphology and dynamics of region R1 (see panel 3 of
Fig. \ref{fig:cloud-evolution}), the other regions have quite similar
characteristics.

Finally, it is worth mentioning that magnetic fields may play only a minor role in the evolution of \hii\ regions since the ram pressure dominates the dynamics of the ionised gas, as was found by \citet{Arthur+11}. However, magnetic fields 
are probably important for the dense gas surrounding \hii\ regions,
but this investigation is beyond the scope of this study 
\citep[see, e.g.,][]{KF19}.


\begin{figure}
	\includegraphics[width=1.0\hsize]{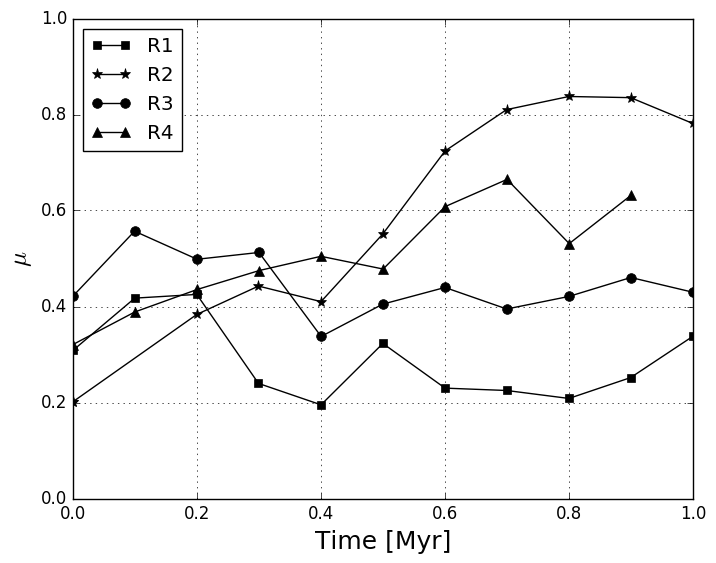}
	\caption{Evolution of the ratio of minimum and maximum axes ($\mu$) for each region. Lower values of this ratio mean regions highly anisotropic, while values tending to one correspond to more symmetric regions.}
	\label{fig:anisotropy}
\end{figure}

\begin{table}
\tiny\scriptsize 
\caption{Properties of the analysed \hii\ regions (see panel 3 of Fig. \ref{fig:cloud-evolution}).
Here, $t_0$ is the time at which each massive star, of mass $M_*$, starts radiating; $R$, $\langle n \rangle$, $\MI$, and $\vrms$ are the radius, mean density, ionised gas mass, and velocity dispersion of each \hii\ region after $\sim 1 \, \Myr$ of evolution.}
\label{table:hii-prop}
\begin{threeparttable}
\begin{tabular}{ccccccc}
\toprule
Region &$M_*$\tnote{a} &$t_0$   & $R$ & \prom{n} & $\MI$   &$\vrms$  \\
name   &$(\Msun)$&($\Myr$) &(pc)&($\ppcc$) &($\Msun$)&($\kms$) \\
\hline
R1     &  15.2   &12.8    &14.5 & 0.9      & 358.5   & 11.2   \\[2pt]
R2     &  10.1   &13.1    & 3.2 & 5.3      & 23.6    & 14.8   \\[2pt]
R3     &  9.9    &13.4    & 3.8 & 2.5      & 17.4    & 16.0   \\[2pt]
R4     &  12.7   &14.3    & 3.7 & 8.5      & 38.1    & 12.4   \\
\bottomrule
\end{tabular}
\begin{tablenotes}\footnotesize
\small
 \item[a] Note that a once massive stars reach $8 \, \Msun$ it starts to radiate. However, the radiation does not stop the mass accretion immediately and the mass of the sink/massive star continue growing.
\end{tablenotes} 
\end{threeparttable}
\end{table}

\begin{figure*}
\includegraphics[width=1.0\hsize]{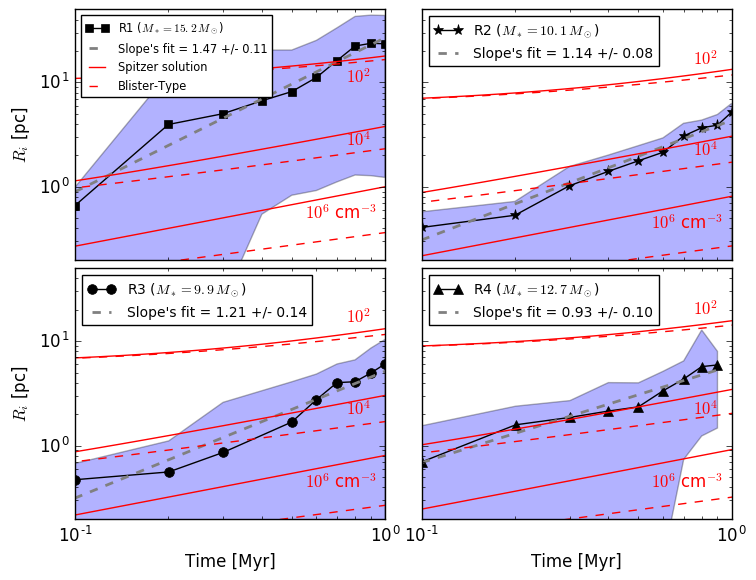}
\caption{Time evolution of the radii of the four \hii\
regions marked in panel 3 of Fig.\ \ref{fig:cloud-evolution}. Grey
dashed lines represent the best power-law fit (the slopes are in the
figure label). The solid red lines correspond to the classical Spitzer
solution (eq.\ (\ref{rII}) in the Appendix) for expanding \hii\
regions immersed in a uniform medium of density
$10^6$, $10^4$, and $10^2 \, \ppcc$, respectively, whereas the
dashed red lines trace the blister-type solution 
(Eq.\ (9) in FST94; see also eq.\ (\ref{ri}) in the Appendix),
assuming the same cloud densities that in the Spitzer solution. Finally, The
blue-shaded region represents the size range between the extremes of the
distance between the star and the ionisation front
for each region.}  \label{fig:rad}
\end{figure*}

\subsection{Evolution of the \hii\ regions}
\label{subsec:hii-regions}

We define an \hii\ region around a massive star as the
connected region around the ionising source where the ionisation fraction is
greater than 0.5. We then calculate the
evolutionary properties of individual \hii\ regions as long as they
remain isolated and powered by a single massive star, which
corresponds to roughly 1 Myr. After this time,
the regions merge with each other or another massive star
appears. Thus, we study four \hii\ regions, which are marked in panel 3
of Fig. \ref{fig:cloud-evolution} as R1, R2, R3, and R4. In Table
\ref{table:hii-prop} we list the main properties of these regions
after $\sim 1$ Myr of evolution.

Here, we focus our discussion in the size, density, and mass of our
selected regions, since other physical properties, such as
temperature, velocity dispersion, and rms magnetic field
strength remain
roughly constant around the mean values of $10^4$ K, $12 \, \kms$, and
$1 \, \mu$G, respectively, for all regions.

We also quantify the anisotropy level by computing the size of the three principal axis. First we calculate the inertia matrix $I_{i,j}=\sum x_i x_j \Delta m$, where $x_i$ and $x_j$ are the coordinates to every cell belonging to the \hii\ region. We then calculate the eigenvectors of the inertia matrix, $I_i$, which represent the inertia momentum. The principal axes can be calculated as $\mu_i = \sqrt{I_i/M_i}$, where $M_i$ is the ionised mass. In Fig. \ref{fig:anisotropy} we plot the ratio of minimum and maximum axes ($\mu =\mu_3)/\mu_1$), and we can see that in general our regions grow anisotropically (with $\mu \lesssim 0.6$), except for the second \hii\ region (star symbols), which tends to be more symmetric after $\sim0.5$ Myr of evolution.

\subsubsection{Size} \label{sec:size}

With the above definition of an \hii\ region, 
and in order to estimate a characteristic size of the \hii\ regions,
we first calculate the total volume ($V$) and then
calculate the radius as $\ri=V^{1/3}$. In Fig. \ref{fig:rad} we
show the evolution of the radius of our four \hii\ regions. We have
chosen the zero of time as the 
moment when each massive star starts to radiate ($t_0$ in Table
\ref{table:hii-prop}).

In the same figure, we have over-plotted the Spitzer solution as well
as the blister-type (FST94) 
estimation for the radius evolution of
blister-type HII regions (solid and dashed red
lines, respectively; see Appendix \ref{ap:models}). Note that both the Spitzer
and the FST94 
solutions describe only the growth of the
cavity within the dense cloud, while our calculations include the
ionised gas that escapes into the diffuse intercloud medium. Thus, the
analytic solutions should be considered only as guidelines. The actual
flow is more similar to the ``champagne'' flow described by
\citet{FTB90}, although without a unique expansion law because our
cloud is filamentary and highly irregular rather than exhibiting a
smooth stratification as assumed by those authors.

Also, in Fig.\ \ref{fig:rad}, the blue-shaded regions
represent
the size range between the minimum and maximum values of the
distance between the star and the ionisation front
for the four \hii\ regions studied.
It is worth noting that the minimum
values of this star-front distance 
occur in the direction toward
the densest parts of the cloud (see, for example, panel 1 in
Fig. \ref{fig:slc_x}). In this direction, the ionisation front
remains stationary, and very close to the star ($\ri \la 0.1$ pc) for
most of the time interval considered in three out of the four \hii\
regions we analysed. This imply that the simulated \hii\ regions 
grow anisotropically, and thus the stars appear
off-centre of the regions during their evolution.

In Fig.\ \ref{fig:rad} we also show grey dashed lines that represent
the best power-law fit to the mean radius of the regions, $\ri$, as a
function of time.  We find that
$\ri \propto t^{1.18 \pm
0.17}$, where the uncertainty range represents the standard deviation among
our four regions. This implies that the expansion velocity and
acceleration are proportional to $\sim t^{0.2}$ and $\sim t^{-0.8}$,
respectively. Thus, once an ionised region
breaks out from its host filament, a blister-type flow is
generated which expands in an
accelerated way towards the low density parts of the MC and the
diffuse medium. This accelerated behaviour has been reported by
\citet{FTB90} and \citet{Arthur+06} for \hii\ regions expanding in a
stratified medium.

\subsubsection{Density}

The left panel of Fig.\ \ref{fig:hii-n-M}
shows the mean density, $\langle n_i \rangle $, as a function of time of the ionised
gas. The initial density is high $\gtrsim 10^{3-4} \, \ppcc$,
but the \hii\ regions deflate quickly as they reach (and
expand through) the WNM, lowering their density by roughly three
orders of magnitude after $\sim$1 Myr of evolution. The final
density for all the regions tends to saturate around $\sim 1-5 \,
\ppcc$.

\subsubsection{Mass}

\begin{figure*}
\begin{multicols}{2}
    \includegraphics[width=\linewidth]{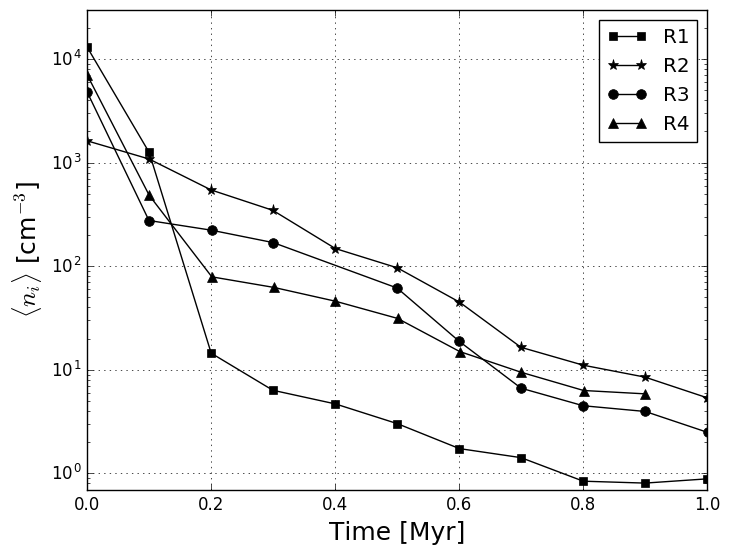}\par 
    \includegraphics[width=\linewidth]{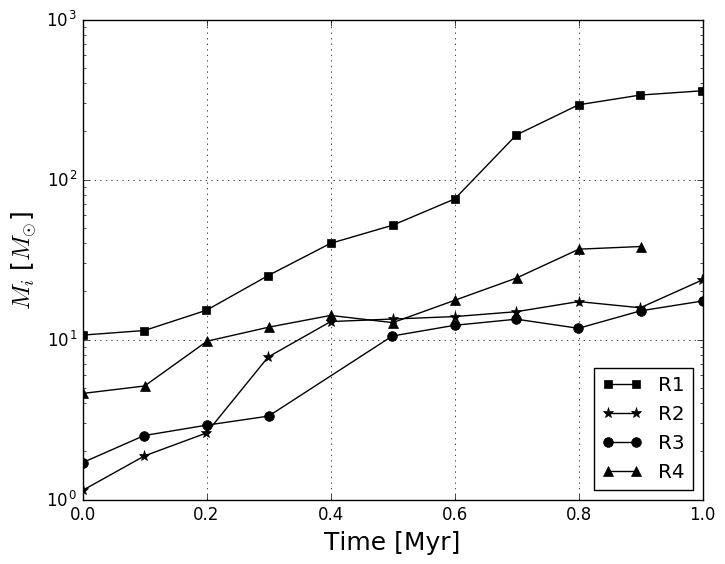}\par 
    \end{multicols}
\caption{Time evolution of the mean number density of ionised gas (left panel) and ionised 
gas mass (right panel) for the four \hii\ regions.} 
\label{fig:hii-n-M}
\end{figure*}

As expected, the ionised mass, $M_i$, of the
\hii\ regions depends on the mass and luminosity of the ionising star (see Fig. \ref{fig:hii-n-M}; right
panel). Region R1 is produced by the most
massive star in our sample 
($15.2 \, \Msun$; see Table \ref{table:hii-prop}),
and it ionises a
gas mass of $~400 \, \Msun$, whereas the stars with masses around $~10
\, \Msun$ (regions R2 and R3) only ionise $\sim 20 \, \Msun$ after $\sim
1 \, \Myr$. 

\subsubsection{The size-density relation}

\begin{figure}
	\includegraphics[width=1.0\hsize]{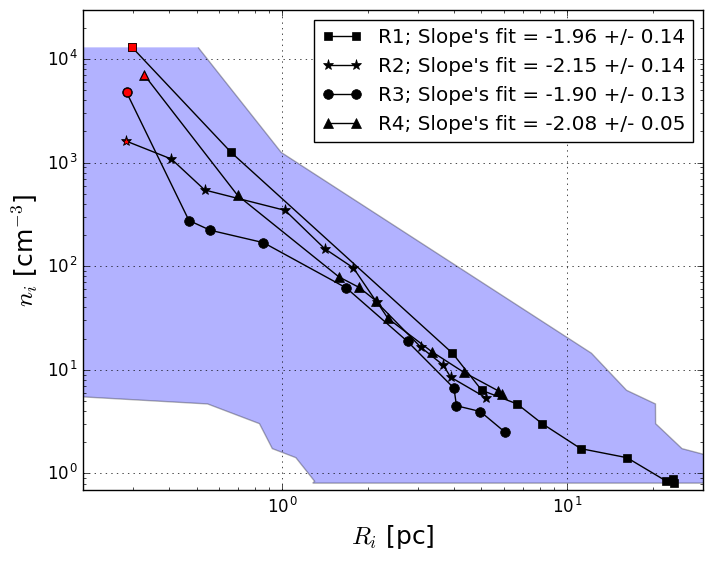}
	\caption{Size-density diagram for the four simulated \hii\ Regions. The blue 
	shadow brackets the radius extremes for the region R1 at each time (see 
	upper-left panel of Fig. \ref{fig:rad}). The average of the 
	slopes is 2.0, i.e., $n_i \propto \ri^{-2}$. The red symbols correspond to 
	the time at which each \hii\ region turns on (at higher density and smaller sizes).}
	\label{fig:size-dens}
\end{figure}

As mentioned in the introduction, \hii\ regions observationally show a
clear correlation between size ($\ri$) and the electronic density
($n_i$) of the form $n_i \propto \ri^{p}$, with $p \simeq -1$, over a
wide dynamical range \citep[see, e.g.,][]{Hunt+09}, although with a
considerable dispersion of $\sim 2$ orders of magnitude. Figure
\ref{fig:size-dens} shows our \hii\ regions in this size-density
diagram. We find a tight correlation, although with a steeper
slope, $p \simeq -2$, for all our \hii\ regions throughout their
evolution.
Some possible
explanations for this discrepancy in the slope are the lack of other
feedback mechanisms (winds and supernova) in our models, or the fact
that our \hii\ regions are produced for a single massive star,
whereas probably most of the regions in the \cite{Hunt+09} sample contain more
than one OB star. 

Note also that, although both the evolution of the size and
density depend on the mass of the ionising star, the
density-size relations do not. Indeed, all
the \hii\ regions are seen to occupy the same 
region in this diagram (see
Fig. \ref{fig:size-dens}). This suggests that the position of a 
given \hii\ region in this diagram can be explained in terms of
its evolutionary state alone. 
On the other hand, this relation also implies that the
mass of the \hii\ regions roughly depends linearly
on their size; that is, $M_i
\propto \ri$, assuming constant density.

\begin{figure*}
     \includegraphics[width=18cm, height=13cm,keepaspectratio]{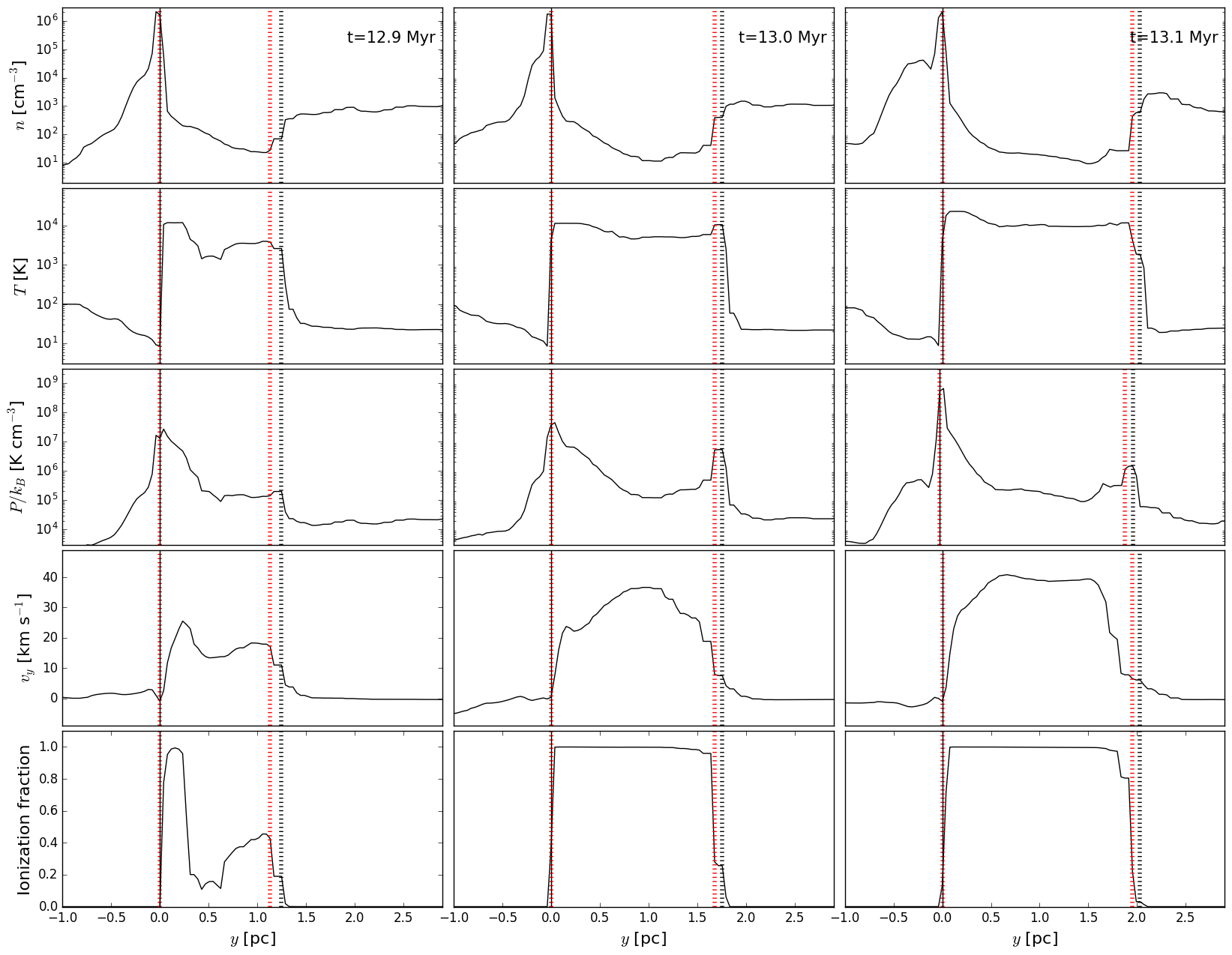}
     \caption{Density, temperature, pressure,
		velocity, and ionisation fraction profiles (from top to
		bottom) for the R1 in the $y$-axis at times 12.9, 13,
		and 13.1 Myrs (left, middle, and right panels,
		respectively). The zero point corresponds to the
		position of the massive star
		(vertical thin black line). The dotted black and red
		vertical lines represent the shock and ionisation
		fronts, respectively.} 
    \label{fig:profiles}
\end{figure*}


\section{Structure of the \hii\ regions} \label{sec:structure}

Our simulations, including a physically reasonable treatment of the radiative
transfer, provide an ideal means for investigating the structure of the
\hii\ regions and their boundaries, especially since there exist
various models in the literature for the expansion of these regions
which are based on different assumptions of their boundary conditions
and internal structure. 

The expansion of \hii\ regions into their parent MCs depends on
whether they are fully embedded in the MC, constituting the
``canonical'' \hii\ regions \citep[e.g.,] [] {Spitzer78}, or break out of the dense clouds and begin expanding into the
diffuse medium, constituting the so-called ``blister'' type of \hii\
regions (e.g., \citet{Whitworth79, FTB90}; FST94).
As summarised in
Appendix A, during their dynamical expansion phase, canonical \hii\
regions are overpressured relative to the 
cloud material, and are
therefore bounded by a shock front, followed downstream by an
ionisation front, with a shocked, dense layer in between the two
fronts. Furthermore, as a consequence of the overpressure in the
region and the dynamical expansion it produces, which reduces the
region's internal density, the region follows an expansion law of the
form $R \sim t^{4/7}$.

Instead, blister-type regions are characterised by a loss of pressure
by means of the ``release valve'' provided by the discharge to the
diffuse medium.\footnote{This is analogous to standard thermal stellar winds or jet streams (de Laval nozzle), which get accelerated to supersonic velocities at the critical breakout point (the nozzle "throat") if the pressure is not balanced.} Therefore, FST94 
argued that these regions are essentially at the same pressure as
their parent cloud and thus do not produce a shock front upstream of
the ionisation front. This in turn implies that the region is bounded,
on its interface with the cloud, by a single ionisation front, with no
preceding shock front and no compressed neutral layer upstream of it.
Also, because in this case the region is not overpressured, its growth
is controlled only by the rate at which the mass inflow through the
ionisation front balances the mass loss to the diffuse medium
(see Fig.\ 1 of FST94), resulting in a slower expansion
rate into the cloud of the form $R \sim t^{2/5}$. 

This conclusion, together with the associated mass photo-\-evaporation
rate of the cloud predicted by FST94 has been questioned by
\citet{Matzner+15}, who argue that FST94 ``incorrectly associate
swept-up mass with ionised mass.'' This is because, according to
\citet{Matzner02}, most of the mass in the \hii\ region is in the
shocked, compressed layer upstream of the ionisation front.
Moreover, the criticism by \citet{Matzner+15} indirectly affects
the accuracy of a model we have presented in previous papers for the
collapse of MCs and their SF activity \citep{ZA+12, ZA-VS14}, which
was based on the prescription by FST94. However, this criticism only
applies if blister \hii\ regions develop a shock at the interface with
the MC. If no shock is present, then there is no swept-up, dense
layer, and so all the material in the region is indeed ionised. 

Our results, described in the previous sections, allow us to
investigate what is the actual nature of the boundary between the
\hii\ region and the cloud, since the radiative transfer and the
cooling rates are followed self-consistently, and thus the ionisation
fraction and the gas structure are realistic, including the presence
and position of the ionisation front and the shock. In Figs.\
\ref{fig:slc_x} and \ref{fig:slc_z} we show the contours where the
ionisation fraction $\chi = 0.5$ (purple) and $\chi =0.9$ (yellow) in
region R1. In the top panels these contours are overlaid onto images
of the density, while in the second and third rows they are overlaid
onto images of the temperature and the pressure, respectively.

The signature of a shock ahead of the ionisation front should be a
region of high density and pressure between the ionisation front and the
ionised region, with a higher temperature than that of the cloud. 
In order to detect shocks in our simulated \hii\ regions, we plot in
Fig. \ref{fig:profiles} profiles of the density, temperature, pressure, velocity ($y$-component), 
and ionization fraction for the R1 and for three different times in the frame of reference
of the massive star. No evidence of a shock upstream of the
ionization front is seen at the edges of the R1 region where it meets
with the densest gas, unless it is too weak to be detected.\footnote{
	Note that we have enough resolution to detect a shock. 
	In the context of stellar winds, for a moderate shock, 
	with a shock velocity $v_s \le 80 \, \kms$, 
	the shock width can be approximated 
	by $1.32 v_{s,100}^{-4.51} n_{{\rm pre},100}^{-1}$ AU \citep{Hartigan+87,RicardoPhDT02},
	where $v_{s,100}$ is the shock velocity in units of $100 \, \kms$ and $n_{{\rm pre},100}$ 
	the preshock density in units of $100 \, \ppcc$ \citep[see also ][]{Gonzalez+04}.
	According to this approximation and using typical values of $v_s=10 \, \kms$ and 
	$n_{\rm pre}=10^3 \, \ppcc$ we expect a shock width of ~0.02 pc, which 
	is comparable to our resolution.} 
Note that this behaviour (a very weak shock toward the densest part) is also 
reported by \citet{Arthur+06} in a stratified density field.

On the other hand, we detect a clear shock toward the general cloud\footnote{
I.e., toward the rest of the cloud, except dense region where the massive star was born.} (marked with 
vertical dotted black lines in Fig. \ref{fig:profiles}) ahead of the ionisation front 
(right vertical dotted red lines) due to the hydrodynamical expansion of the \hii\ region.
Thus, our \hii\ regions expand in a hybrid way: towards the general cloud we detect 
a moderate shock, whereas we do not detect any shock signature (or it is too weak to be 
detected) toward the dense clump.

Furthermore, this region also exhibits features of the
blister type, as it has ample sections where it connects directly to
the warm, diffuse gas. In these sections, it expands in an
accelerated manner, in qualitative agreement with the analytical
predictions of \citet{FTB90}, although, due to the non-unique density
stratification, our region does not exhibit a unique acceleration law. 
Thus, we conclude that neither the classical $t^{4/7}$ nor the FST94 $t^{2/5}$
expansion laws apply to our regions, which instead expand in an
accelerated way as $\ri \sim t^{1.2}$, in qualitative agreement with \citet{FTB90}. 

\section{Discussion}\label{sec:discusion}

\subsection{Implications of the size-density relation}

\citet{Hunt+09} interpreted the observed size-density relation 
for \hii\ regions ($n_i
\propto \ri^{-1}$) as being a consequence of the observed
Larson relation for molecular clouds
($n_{H_2} \propto R_{H_2}^{-1}$), suggesting that
\hii\ regions retain an imprint from the molecular
environment in which they form. However, this Larson relation 
has been repeatedly questioned, since it
may be the result of various selection effects,
rather than  real property of MCs
\citep[e.g.,][]{Kegel89,Scalo90,VS+97,BP-MacLow02, BP+12}.
Indeed, numerical simulations of MC formation and evolution do not
show any evidence for the appearance of such a density-size scaling
appearing in the clouds unless a selection criterion of roughly
constant column density is imposed \citep{BP-MacLow02, Camacho+16}.
On the other hand, our
results strongly suggest that individual \hii\ regions do follow
an evolutionary path in the $n_i-\ri$ diagram. This
evidence suggests that, the size-density scaling of \hii\
regions may be a real effect, although, contrary to the interpretation
of \citet{Hunt+09}, unrelated to the density structure of the parent MC.

\subsection{Comparison with previous results}

Analytical models of blister-type expansion of \hii\ regions are
based on different hypotheses. While FST94
assume that the ionization
front advances with no shock towards the dense region of the cloud,
\cite{Matzner02} assumes that it is through this shock that the \hii\
region incorporates most of its mass (see
App. \ref{ap:models}). However, in the
present study we have shown that both of these hypotheses are
satisfied at the same time in different parts of
the \hii\ regions. The assumption by FST94 is valid
toward the dense gas (where the massive star was born),
while the assumption of \cite{Matzner02} can be applied toward the
rest of the cloud. On the other hand, the \hii\ regions we analysed
expand in an accelerated way toward the low density gas (WNM) as
predicted by \cite{FTB90} 
in a stratified medium \citep[see also][]{Arthur+06}.

Although the structure and dynamics of massive star-forming
regions are quite complicated and far from the idealised geometries
assumed in analytical models, an approximate
comparison is possible.  The size-density relation ($n_i \propto
\ri^{-2.0}$) implies that the ionised gas mass, $M_i
\propto \ri$, or equivalently, $\dot{M}_i \propto \dot{\ri}$. As we
found in \S \ref{subsec:hii-regions}, the \hii\ region radius depends
roughly on time as $\ri \propto t^{1.19}$. Therefore, the
mass ionisation rate is $\dot{M}_i \propto
t^{0.19}$. Note that this dependence is quite close to $\dot{M}_i
\propto t^{1/5}$ predicted by FST94 (Eq. \ref{eq:dotm}) for blister
type \hii\ regions. In turn, this implies that the
blister-\hii\ region approximation used in the analytical model by
\citet{ZA+12} is justified.

On the numerical side, several works have included ionising
feedback \citep[e.g.,][]{Dale+05,Colin+13,Geen+15}, although they only
study global properties of MCs, rather than the internal structure of 
the \hii\ regions. On pc scales, \citet{Arthur+11} studied
the evolution of individual \hii\ regions, albeit only during the 
embedded phase, so we cannot compare directly our findings with 
these works. In any case, they showed that the presence of a 
magnetic field does not modify the \hii\ region structure in a 
fundamental way. Finally, although we have investigated only one 
simulation, our results can be considered robust since our simulated
MC is highly chaotic, due to the turbulence injected in a
self-consistent way, and therefore the idealised initial conditions 
are quickly forgotten. Moreover, we have investigated a sample of four 
\hii\ regions that span a range of behaviours, from which we have 
extracted some meaningful averages.

\subsection{Limitations}

As we are interested in studying the effect of the ionising feedback,
in this study we have neglected other feedback
processes, such as winds, supernova explosions and radiation
pressure, which can help disperse the cloud and further
reduce both the SFR and the SFE. However, these feedback effects can
also enhance the positive feedback through the injection of
mechanical energy, and can possibly bring the
slope of the size-density relation of our \hii\
regions closer to the observed value. Thus, the net effect of
these feedback mechanisms is so far unclear, and
more numerical research is necessary in order to study their relative
importance \citep[see, e.g.,][]{Bastian+16,Wareing+17}.

Also, our spatial and temporal output
resolutions\footnote{Corresponding to 0.03 pc and 0.1 Myr,
respectively. The temporal resolution refers to the time
interval between output
snapshots.} are insufficient to resolve the stages of
hyper-compact/compact \hii\ regions.\footnote{The typical sizes and life
times of compact \hii\ regions are $\leq 0.2$ pc and $\sim$0.3 Myr,
respectively \citep[][and references therein]{Garay-Lizano99}.} 
Particularly, we do not resolve the initial
Str\"omgren radius in our star formation regions.
However, this is not an issue since we are interested in studying the
expansion of \hii\ regions at the scale of the MCs, as a means of destroying
them. Although our numerical simulations include magnetic
fields, it is still necessary to assess
its detailed influence on the feedback processes through
comparison with non-magnetic simulations, a task we defer to a future
contribution.

Finally, the FLASH radiative transport module does not take into account the absorption of UV photons by dust grains inside the \hii\ regions. Also, models by \citet{Hosokawa+09} suggest that the ZAMS model we use can overestimate the ionising luminosity. Both effects tend to overestimate the number of UV photons emitted by a massive star, so properties of \hii\ regions such as size should be taken as upper limits \citep[see][for a detailed discussion]{Peters+10}. However, the power law index of the radius growth should remain unchanged, as well as their associated relations.
In addition, the effects of molecular hydrogen destruction by
photodissociation from massive and intermediate-mass stars is not
considered here, and, as discussed by \citet{Diaz-Miller+98}, this
effect makes the cloud evaporation process more effective.

\section{Summary and Conclusions} \label{sec:conclusions}

In this paper, we have studied the evolution of the physical
properties of \hii\ regions in radiation-MHD simulations of MCs formed
by diffuse converging flows in the presence of magnetic fields and
massive-star ionisation feedback.

Our simulations reproduce the rich morphology observed around
\hii\ regions, such as elephant trunks,
cloudlets, champagne flows, etc. Due to the 
highly complex structure of our filamentary clouds, the \hii\ regions
grow anisotropically, causing the massive stars to appear off-centre of the ionised
regions. Our simulated \hii\ regions expand
in a hybrid way in our filamentary
clouds: 
towards the intermediate-density regions in the
cloud, the \hii\ regions expand according to the classical theory,
developing a shock ahead of the ionisation front, while towards the
densest parts of the cloud, no shock is apparent, and the ionisation
front stalls in 3 out of 4 cases, and advances according to classical
theory in the other case. Finally, towards the diffuse, warm gas, the
regions expand roughly according to the ``blister'' case, at an
accelerated pace. On average, the radius of the regions is dominated by
this latter mode, so that the average radius grows at an accelerated
pace, roughly as $\ri \sim t^{1.2}$.

Our \hii\ regions exhibit a tight
relation between size and average density, $\langle
n_i \rangle$, of the form $\langle n_i \rangle \propto \ri^{-2}$. This
implies that, on average, the mass ionisation
rate is $\dot{M}_i \propto t^{0.19}$, which is in good agreement with
the analytical prediction by FST94 ($\dot{M}_i
\propto t^{0.2}$). Therefore, we conclude that the analytic
prescription for the rate of mass ionisation used in the model by
\citet{ZA+12} is adequate, contrary to
\citet{Matzner+15}.  Interestingly, the electron
density-size relation we observe in our \hii\ regions is the
result of the expansion mechanism itself, rather than an imprint of a
Larson-type relation in their environment, which, in addition, is not
observed in general in this kind of simulations.


\section*{Acknowledgements}
We thank the anonymous referee for helpful comments and suggestions, which helped to improve this manuscript.
We also gratefully acknowledge useful comments from Guillermo Tenorio-Tagle.
MZA and EVS acknowledge financial support from CONACYT grant number
255295 to EVS.  RG acknowledges UNAM-PAPIIT grant number IN112718.
The research of LH was supported in part by NASA grant NNX16AB46G.
JBP acknowledges UNAM-PAPIIT grant number IN110816.
The visualization was carried out with the {\tt yt} software \citep{yt}. 
The FLASH code used in this work was in part developed by the DOE 
NNSA-ASC OASCR Flash Center at the University of Chicago. The authors thankfully acknowledge computer resources, technical advise and support provided by {\it Laboratorio Nacional de Superc\'omputo del Sureste de M\'exico} (LNS), a member of the CONACYT network of national laboratories.

\appendix

\section{Dynamical evolution of \hii\ regions}\label{ap:models}

The different phases of the evolution of an \hii\ region
have been studied in several previous works \citep[see, e.g.,][]{Spitzer78,Whitworth79,Dyson} 
At first, a \hii\ region expands until the ionisation balance
is reached, when the ionisation front is located at the initial
Str{\"o}mgren radius $R_{0}$. Afterwards, 
the \hii\ region
enters into a second stage of dynamical evolution as long as
the ionised gas has a higher pressure than the neutral ambient
medium. In the particular case of a massive star located near the
surface of a molecular cloud, the \hii\ region is radiation-bounded
on the inner part of the cloud, and density-bounded on the outer
part \citep[e.g.,][]{Whitworth79}.
Consequently, a cometary \hii\ region
is produced, in which the ionised gas flows into the low-density
medium.

FST94 estimated
the maximum number of massive stars that can form within a molecular
cloud. These authors pointed out that the most efficient destruction
mechanism is the evaporation of the cloud by stars located near
the cloud's boundary. In that case, the growth of a \hii\ region inside
the cloud is due to the mass flux ({\it a blister-type} mass loss)
that expands into the external low-density medium $n_0$ at a velocity
equal to the sound speed $c_I$ in the \hii\ region. In their model,
it is assumed that the mass loss into the environment is equal to
the mass gained by the expansion of the \hii\ region inside the cloud,
that is,

\begin{eqnarray}
\pi R_{i}^2 m_{p} \langle n_i \rangle c_{s} = 2 \pi \ri^2 m_{p} n_0 \dot R_{i}\,,
\label{mgml}  
\end{eqnarray}

\noindent
being $R_{i}$ is the position of the ionisation front, \prom{n_i} the mean
density of the ionised gas, $c_{s}$ the initial expansion speed of the
ionised gas into the external low-density medium (which is assumed equal to 
the sound speed in the ionised gas), and $m_{p}$ the proton mass.

Consequently, the velocity of the ionisation front into the cloud is
given by,

\begin{eqnarray}
\dot R_{i} \simeq {{\langle n_i \rangle}\over{2\,n_0}}\,c_{s}\,,
\label{dri}  
\end{eqnarray}

Considering that the total ionised mass remains constant, it follows
that the position of the ionisation front at a time $t$ within the cloud
is obtained by,

\begin{eqnarray}
R_{i} \simeq R_0\,\biggl(1 + {{5}\over{2}}\,
 {{c_s\,t}\over{R_0}}\biggr)^{2/5}\,.
\label{ri}  
\end{eqnarray}

\noindent
For this, these authors neglected the effects of the weak shock
of the expanding \hii\ region due to the mass loss from the blister,
and pointed out that both the mass and ionisation balance determines
the evolution.

Assuming a constant luminosity during the main-sequence stage of the
star, the cloud evaporation rate induced by a single star
located near the cloud's boundary calculated by these authors is given by,

\begin{eqnarray} \label{eq:dotm}
\dot M_{i} (t) \simeq 2\pi\,R_0^2 m_p c_s n_0\,\biggl(1 + {{5}\over{2}}\,
 {{c_s\,t}\over{R_0}}\biggr)^{1/5}\, .
\label{ifpos}  
\end{eqnarray}

\noindent

On the other hand,
\citet{Matzner+15} argued that this photoevaporation
rate incorrectly associates the swept-up mass with ionised mass. In another work, \citet{Matzner02} presented a simple treatment of 
the momentum generation
by an \hii\ region. This author takes into account the inertia of the dense shell
produced by the shock moving into the cloud. In this stage of evolution
($\ri > R_0$), the expansion of the ionisation front is caused by the pressure
gradient between the \hii\ region and the environment of neutral gas.
It is assumed by the author that nearly all of the mass originally located
inside the radius $\ri$ remains within the shocked shell.\footnote{It is worth to mention
that this assumption represents an important difference with respect to the model
developed by FST94 
in which the effects of the compressive effects
are neglected due to the mass loss from the blister.}
According to this author, if the ionised gas is effectively isothermal in blister
regions, the ionisation front tends to a D-critical case, for which,\footnote{ 
Hereafter, we follow the notation used in \citet{Matzner02}.}

\begin{eqnarray}
u_{II} - \dot r_{II} = -c_{II}
\label{uII}
\end{eqnarray}

\noindent
where $u_{II}$ is the velocity of the ionised gas relative to the cloud,
$\dot r_{II}$ is the expansion speed of the ionisation front, and $c_{II}$ is
the sound speed of the ionised gas. Consequently, the rate at which mass is ionised
is given by,

\begin{eqnarray}
\dot M_{II} = \rho_{II} (\dot r_{II} - u_{II})\, 2\pi r_{II}^2 \,
\label{MII}
\end{eqnarray}

\noindent
where $\rho_{II}$ is the density within the \hii\ region.

Since the \hii\ region expands supersonically with respect to the molecular
gas, it is bounded by a thin shocked layer that is located near the ionisation
front. Assuming a radial expansion, 
\citet{Matzner02} found a self-similar
solution for $r_{II} \gg R_{\tiny \mbox{St,0}}$ (with $R_{\tiny {\mbox{St,0}}}$ the
initial Str$\ddot{\mbox{o}}$mgren radius) given by,

\begin{eqnarray}
r_{II} = {{49} \over {6}}\,\,R_{{\tiny \mbox{St,0}}}^{3/7}\,(c_{II} t)^{4/7}
\label{rII}
\end{eqnarray}

\noindent
Considering the momentum of the radial motion of the expanding shell,
the mass evaporated for a blister region can be estimated by,

\begin{eqnarray} \label{mdest}
\delta M_{\tiny{\mbox dest}}= 1.2 \times 10^4 \,
 \biggl({t \over {3.7\,{\mbox Myr}}}\biggr)^{9/7}
 \biggl({{N_{\tiny{\mbox H,22}}} \over 1.5}\biggr)^{-3/14}\\
\nonumber
\times M_{\tiny \mbox {cl,6}}^{1/14} \, S_{49}^{4/7} {\mbox M_{\odot}} ,
\end{eqnarray}

\noindent
where $N_{\tiny{\mbox H,22}}$ is the mean hydrogen column density in
units of $10^{22}$ cm$^{-2}$, $M_{\tiny \mbox {cl,6}}$ is the mass of the
cloud in units of $10^6$ M$_{\odot}$, and $S_{49}$ is the ionising
photons' rate in units of $10^{49}$ photons per second. According to this author, equation [\ref{mdest}]
for the mass evaporated agrees with predictions by 
\citet{Williams+97} who argued that only 10$\%$ of the mass of a GMC becomes
stellar, within 1$\%$.

%

\bibliographystyle{mnras}
\bibliography{references}

\bsp	
\label{lastpage}
\end{document}